# Iterated Gain-based Stochastic Filters for Dynamic System Identification: Annealing-type Iterations and the Filter Bank

Tara Raveendran[1], Debasish Roy[2,*] and Ram Mohan Vasu[1]

[1]Department of Instrumentation and Applied Physics, Indian Institute of Science, Bangalore

[2]Computational Mechanics Lab, Department of Civil Engineering, Indian Institute of Science, Bangalore

[*]Corresponding Author; Email: royd@civil.iisc.ernet.in

*Abstract*

*A novel form of nonlinear stochastic filtering employing an annealing-type iterative update scheme, aided by the introduction of an artificial diffusion parameter and based on the Gaussian sum approximations of the prior and posterior densities, is presented. The proposed Monte Carlo filter bank conforms in structure to the parent nonlinear filtering (Kushner-Stratonovich) equation, as reflected in the additive gain-based updates, and possesses excellent mixing properties enabling better explorations of the phase space of the state vector. The performance of the filter bank, presently assessed against a few carefully chosen numerical examples, provide ample evidence of its substantively improved performance in terms of filter convergence and estimation accuracy vis-à-vis a few other competing filters especially in higher dimensional dynamic system identification problems including cases that may demand estimating relatively minor variations in the parameter values from their reference states.*

**Keywords:** Stochastic filtering; iterated gain; ensemble square root filter; Gaussian sum approximation; Kushner-Stratonovich equation; dynamic system identification; parameter estimation

## 1. Introduction

Dynamic system identification aims at estimating the hidden state processes that solve the system or process model, often in the form of stochastic ordinary differential equations (SDEs), given a set of noisy partial observations, which are typically characterized by the observation SDEs whose drift fields are known functions of the system (process) states. The 'estimate of a state' often stands for its mean (first moment) with respect to the probability density function (PDF) of the instantaneous state conditioned on the observation history till the current time. Bayesian filtering, which is a computationally feasible and popularly adopted route in obtaining the filtering PDF, involves a two-step recursive procedure consisting of the prediction and update stages. While the prediction stage requires recursively propagating the process or system model in time, the predicted solution is modified in the update stage in order to assimilate the currently available observation consistent with a recursive form of the generalized Bayes' formula [1] and thus characterize (marginals of) the filtering PDF. The Kalman filter has been a major breakthrough [2], providing for an analytical scheme to arrive at the exact posterior PDF for a linear Gaussian dynamic state space model. Nonlinear dynamical systems with non-Gaussian additive/multiplicative noises may also be dealt with, albeit sub-optimally, with extended Kalman filters (EKFs) that employ linear approximations to the signal-observation dynamics. But the EKF and its variants [3] may perform quite poorly where the dynamics are significantly nonlinear due to the imprecise Gaussian approximation of the transition law of the signal-observation process.

Particle filters (PFs) rely on a first order Markov model for the time-discretized signal-observation processes and implement a recursive Bayesian filter by Monte Carlo (MC) simulations [4]. Over a given time-step, they use particles, which are independently sampled and weighted realizations of the random variables (representing the instantaneous filtered states) to approximate the continuous filtering PDF by random (empirical) measures where the weights obtain the likelihood of the current observation given the particles. Being free from the approximations involving linearizations, PFs are endowed with the universality that have seen their applications to the identification of a wide-ranging array of noisy nonlinear dynamical systems encountered in target tracking, digital communications, chemical engineering etc. [5, 6, 7]. Efforts to use a form of analyticity characteristic of the Kalman filter within the framework of particle filtering have led to the development of a semi-analytical PF [8]. Such PFs transform the nonlinear system/observations to an ensemble of piecewise linearized equations so that Kalman filter can be used for each linearized system to yield a family of conditionally Gaussian posterior PDFs whose weighted sum yield the filtering PDF. The accruing advantage of reduced sampling variance however comes at the cost of a substantively increased computational overhead as the current observation must be repetitively assimilated for each linearized system. In any case, PFs generally perform poorly in applications involving higher dimensional process models as the weights tend to collapse to a point mass [9] and the necessary sample size needed to counter such weight degeneracy could be practically unattainable.

The ensemble Kalman filter (EnKF), which may be viewed as an MC version of the Kalman filter, has found applications in higher dimensional filtering problems as in oceanographic and atmospheric modelling [10], where the infeasibility of storage-cum-manipulation of large state error covariance matrices and the need for extensive tuning of the noise covariance matrices have been the primary issues hindering the use of Kalman filter. The EnKF uses an ensemble of system states predicted through the process dynamics, thus avoiding the Gaussian closure through linearization as in the EKF. In addition, it uses a pseudo-ensemble of observed states towards computing the Kalman gain via particle-based approximations to the error covariance matrices which in turn are used in the evaluation of the updated particles. The additive nature of the gain-based updates insures against possible weight collapse (and hence particle impoverishment) often encountered with PFs.

Every filtering algorithm may be viewed as a numerical or analytical scheme to determine, approximately or otherwise, the state estimates (or the filtering PDF) from the parent nonlinear filtering equation, i.e. the Kushner-Stratonovich (KS) equation (or the Zakai equation) [11]. Variants of algorithms for branching particle approximations convergent in distribution to the solutions of the KS equation form another subclass of MC-based methods [12, 13]. By way of a 'maximal' utilization of the current observation, iterative filters (e.g. the Iterative Gain Based Stochastic Filter (IGSF) and the Unscented Iterative Gain Based Stochastic Filter (UIGSF)) have recently been developed by the authors [14] so as to iteratively improve the gain-based update term in the current time step whilst conforming to the form of the nonlinear KS equation [11]. These iterations, aimed at accelerating the filter convergence and performed over a pseudo-time variable $\zeta$, are basically designed to drive the current innovation term to a zero-mean martingale (e.g. a Brownian motion) in $\zeta$. Continuing with the same theme, the purpose of this article is to suggest substantive modifications in the iterative filtering scheme by bringing in an annealing-type parameterization in the update scheme through an artificial diffusion parameter (ADP) whilst approximating the associated PDFs through Gaussian sums. The iterative update stage requires ADP-parameterized repetitive computations of a Kalman-like gain $\mathcal{K}_i^l$ ($i$ being the temporal recursion step and $l$ the iteration index for a fixed $i$), consistent with the nonlinear KS equation, with the initial guess $\mathcal{K}_i^0$ evaluated on similar lines as the ensemble square root filter (EnSRF) [15]. The Gaussian sum filter bank [16] helps exploring the phase space of the state variables better and the added diversity in the particles enables easier adaptation/reconciliation of the process dynamics with the measured variables. The ADP, which may be lowered to zero over successive iterations at a much faster rate (allowing even for a discontinuous scheduling) than is feasible with the conventional simulated annealing, also helps enhance the so called 'mixing property' [17] of the iterative update kernels. This additional layer of parameterization in the iterated update scheme thus enable, in a sense, an optimal assimilation of the measured vector within the estimate at the time instant $t_i$ for every $i$ which is not ensured in standard forms of PFs. An attempt is made to provide adequate numerical evidence of the enhanced filter performance with the introduction of some of these novel elements.

## 2. Statement of the Problem

Let $(\Omega,F,P)$ be a complete probability space with $F_t$, $t \geq 0$, being the $\sigma$-algebra generated by all the noise processes involved in the presentation to follow at a given time $t$. Clearly the size of the set $F_t$ increases with increasing $t$ and the collection of sets $\mathbf{N}_t := \{F_s : 0 \leq s \leq t\}$ defines the so called increasing 'filtration' as $t$ increases. Also the time interval of interest $[0, \tau]$ is discretized as $0 = t_0 < t_1 ..... < t_i ... t_L = \tau$ with $\tau_i = (t_{i-1}, t_i]$, $i \in \mathbf{Z}^+$. The process model describing the evolution of the so-called 'hidden' states of a continuous-time dynamic system subject to random disturbances may often be represented by the Ito stochastic differential equation (SDE) [18]:

$$dX(t) = \mathcal{F}(X(t), t)dt + G(t)dB(t) \tag{2.1}$$

where the state vector $X(t) \in \mathbb{R}^{n_x}$ is a time-continuous signal, $\mathcal{F} : \mathbb{R}^{n_x} \times \mathbb{R} \to \mathbb{R}^{n_x}$ is the system transition function, $B(t) = \{B^{(r)}(t) : r \in [1, q]\}$ is a $q$-dimensional vector of independently evolving zero-mean $F_t$-Brownian motion processes with $B^{(r)}(0) = 0$ and $E\{(B^{(r)}(t) - B^{(r)}(s))^2\} = t - s$, where E denotes the expectation with respect to the measure $P$, and $G : \mathbb{R}^{n_x} \times \mathbb{R} \to \mathbb{R}^{n_x} \times \mathbb{R}^r$ is the diffusion or volatility co-efficient. System identification typically involves estimating the uncertain or inadequately known parameters in the system model and the solution, within the stochastic filtering framework, requires declaring the parameters as additional states denoted as $\mu(t) \in \mathbb{R}^{n_\mu}$. The original state space model (SSM) is thus augmented by allowing $\mu(t)$ to artificially evolve as Brownian motions, as depicted through the following system of zero drift SDEs:

$$d\mu(t) = G_\mu dB_\mu(t) \tag{2.2}$$

where $G_\mu \in \mathbb{R}^{n_\mu \times n_\mu}$ is the diffusion coefficient matrix and $B_\mu(t) \in \mathbb{R}^{n_\mu}$, a zero mean Brownian noise vector process. In fact, restricting Eq. (2.2) over different time sub-intervals $\{(t_i, t_{i+1}], i = 0, 1, ...\}$, $\mu(t)$ may be interpreted as a collection of local Brownian motions (i.e. different mean vectors over different sub-intervals), or, more generally, as local martingales (see [1] for a definition of local martingales). The augmented state vector (with parameters as additional states) is now denoted as $\tilde{X} := [X^T, \mu^T]^T = \{\tilde{X}^{(j)} \mid j \in [1, J]\} \in \mathbb{R}^J; J = n_x + n_\mu$. The response of the dynamic system is partially observed through the noisy and continuous measurement process given by the SDEs (written below in the integral form):

$$Z(t) = \int_0^t \mathcal{A}(\tilde{X}, s)ds + G_z B_z(t) \tag{2.3a}$$

or more appropriately, since the measurements arrive in a time-discrete manner, by an algebraic counterpart of the above equation:

$$Z(t_{i+1}) := Z_{i+1} = \mathcal{H}(\tilde{X}_{i+1}, t_{i+1}) + G_z B_z(t_{i+1}) \tag{2.3b}$$

Here $\tilde{X}_{i+1} = \tilde{X}(t_{i+1})$, $Z = \{Z^{(m)} : m \in [1, d]\} \in \mathbb{R}^d$ denotes the vector of measurements and $G_z B_z \in \mathbb{R}^d$ a $d$-dimensional Brownian motion (measurement noise) with zero mean and covariance matrix $G_z G_z^T \in \mathbb{R}^{d \times d}$. The measurement vector function:

$$\mathcal{H} := \{H^{(k)}(\tilde{X}, t) : \mathbb{R}^{n_x + n_\mu} \times \mathbb{R} \to \mathbb{R}; \ k \in [1, d]\}$$

maps the signal process $\tilde{X}(t)$ to $\mathbb{R}^d$. Let $Z_{1:i} := \{Z_1, ..., Z_i\}^T$ denote the set of measurement vectors till $t = t_i$. The process equations (2.1) and (2.2) may now be combined to yield the nonlinear SSM:

$$d\tilde{X}(t) = \tilde{\mathcal{F}}(\tilde{X}, t)dt + \tilde{G}(t)d\tilde{B}(t) \tag{2.4}$$

where $\tilde{\mathcal{F}} := \{\tilde{\mathcal{F}}^{(j)}, j \in [1, J]\} \in \mathbb{R}^J$ and $\tilde{G}(t) \in \mathbb{R}^{J \times J}$ are respectively the nonlinear drift vector and the diffusion coefficient matrix. The required conditions for the existence of weak solutions [18] to the above SDEs are assumed to be satisfied. The nonlinear system of SDEs (2.4), whose solution yields the predicted response for the filter, may be semi-analytically integrated following a derivative-free and explicit local (piecewise in time) linearization [19] of $\tilde{\mathcal{F}}$ leading to:

$$d\tilde{X}^L(t) = \tilde{\mathcal{Q}}(\tilde{X}_i,t_i)\tilde{X}^L dt + \tilde{G}(t)d\tilde{B}(t), \quad t \in (t_i, t_{i+1}] \tag{2.5}$$

for $t \in (t_i, t_{i+1}]$ with $\tilde{X}^L(t) \to \tilde{X}(t)$ (only in law, implying that weak stochastic solutions are admissible) as $t_{i+1} - t_i \to 0$. In view of the fact that the error due to local linearization may be weakly corrected via a Girsanov change of measure [20,21] and in order to maintain notational simplicity, we will henceforth refer to the linearized (predicted) solution $\tilde{X}^L(t)$ as $\tilde{X}(t)$.

The nonlinear filtering problem is solved by computing the conditional expectation (estimate) $\pi_t(\hat{\tilde{X}}) := E_{\hat{P}}(\hat{\tilde{X}}(t) | N_t^Z) = E_{\hat{P}}(\hat{\tilde{X}}(t) | \{Z(s); 0 < s \leq t\})$ or, more generally, the associated (marginal) distribution $\hat{P}(\hat{\tilde{X}}(t) | N_t^Z)$ (or its density $\hat{p}$, called the filtering or posterior PDF, if it exists) where the filtration at time $t$ generated by the observation $\sigma$-algebra is given by $N_t^Z := \{F_s^Z | s \leq t\}$. Note that, for a given $t$, the filtered random variable drawn from (the numerical approximation to) the posterior distribution $\hat{P}$ is here denoted by $\hat{\tilde{X}}(t)$, which may be distinguished from the predicted process denoted by $\tilde{X}(t)$. It is assumed that the distributions $\hat{P}$ and $P$ are absolutely continuous with respect to each other.

## 3. Filtering Scheme

In order to maintain consistency of the filtering formulation with the KS equation, it is instructive to write the latter in an integral form for $t \in (t_i, t_{i+1}]$ as:

$$\pi_t(\phi) = \pi_{t_i}(\phi) + \int_{t_i}^{t} \pi_s(L_s(\phi))ds + \sum_{m=1}^{d} \int_{t_i}^{t} \{\pi_s(M_s^{(m)}(\phi)) - \pi_s(H^{(m)}(\hat{\tilde{X}}(s),s))\pi_s(\phi)\}dI_s^{(m)}$$

(3.1)

where $\pi_t(\phi(\hat{\tilde{X}}))$ is the estimate of $\phi(\tilde{X})$ at time $t$ and, without a loss of generality, $\phi$ is assumed to be a scalar-valued function for a simpler exposition. Choosing, for instance, a family of such functions $\phi^{(j)}(\tilde{X}) = \tilde{X}^{(j)}$, $1 \leq j \leq J$, one can determine the estimate of (components of) the augmented state vector $\tilde{X}$. $\{I_t^{(m)}(x)\} := \{Z_t^{(m)} - \pi_t(H^{(m)}(x,t))\}$, $x \in \mathbb{R}^J$, is referred to as the innovation vector. The operators $L_t$ and $M_t^{(m)}, m \in [1,d]$, are defined through:

$$L_t(\phi(x)) := \frac{1}{2} \sum_{k=1}^{J} \sum_{j=1}^{J} \tilde{\mathscr{G}}_{kj}(t) \frac{\partial^2 \phi(x)}{\partial x_k \partial x_j} + \sum_{j=1}^{J} \tilde{\mathscr{F}}^{(j)}(x,t) \frac{\partial \phi(x)}{\partial x_j}, \quad x \in \mathbb{R}^J \tag{3.2a}$$

and $M_t^{(m)}(\phi(x)) = H^{(m)}(x,t)\phi(x)$ \hfill (3.2b)

with $\tilde{\mathscr{G}}_{kj}(t) := [\tilde{G}(t)\tilde{G}(t)^T]_{kj}$ \hfill (3.2c)

The mathematical problem of stochastic filtering may be stated as recursively updating the predicted stochastic process $\phi(\tilde{X}(t))$ to the filtered stochastic process $\phi(\hat{\tilde{X}}(t))$ such that $I_t^{(m)}$ is reduced to a zero-mean Brownian motion (or, more generally, a zero-mean martingale) as $t \to \infty$. Consistent with our recent work on iterative stochastic filtering [14], the second term on the RHS of the KS equation (3.1) is approximated as

$$\int_{t_i}^{t} \pi_s(L_s(\phi)) ds \approx \int_{t_i}^{t} \pi_i(L_s(\phi)) ds = \int_{t_i}^{t} E_{\hat{P}}(L_s(\phi) | N_i^Z) ds,$$

where $\pi_i(.) := \pi_{t_i}(.)$ and $N_i^Z := N_{t_i}^Z$. By so uncoupling the prediction and updating stages in the proposed filter over $t \in (t_i, t_{i+1}]$ via this approximation, the first two terms on the RHS of Eq. (3.1) yields the expectation $E_P(\phi(\tilde{X}(t))$ involving the predicted process $\tilde{X}(t)$ that solves Eq. (2.5) and is referred to as Dynkin's formula [18]. Subject to the above approximation, the predicted solution $\phi(\tilde{X}(t))$ in the proposed filter may thus be interpreted as being correspondent to the first two terms on the RHS of the KS equation (3.1). Also, the third term on the RHS involves a sum of integrals over weighted innovations, wherein each scalar innovation $I_t^{(m)}$ is representative of the prediction error if the predicted process $\tilde{X}(t)$ is used to compute the measurement function $H^{(m)}(\tilde{X},t)$. Accordingly, the coefficient weights $k_t^{(m)}(\phi) = \pi_t(M_t^{(m)}(\phi)) - \pi_t(H^{(m)}(\tilde{X}(t),t))\pi_t(\phi)$ may be looked upon as the coefficient gain terms (which, upon integration, build up the gain matrix) that drive $\{I_t^{(m)}\}$ to a set of $d$ zero-mean scalar Brownian motions. However, unlike conventional filtering, the iterative filtering scheme additionally introduces a pseudo-time parameter $\zeta$ and aims at driving the $\zeta$-parameterized innovation process $I_{t,\zeta}^{(m)} := Z_t^{(m)} - \pi_{t,\zeta}(H^{(m)}(\tilde{X}(t,\zeta),t))$ to a zero-mean Brownian motion in $\zeta$, for a fixed $t$ and for each $m$, through iterations (i.e. recursions over $\zeta$). Ideally, one hopes at recovering the true estimate (that satisfies the KS equation at time $t$) as $\pi_t(\phi) = \lim_{\zeta \to \infty} \pi_{t,\zeta}(\phi)$. However, with the current unavailability of a mathematically consistent yet computationally feasible stopping criterion, one typically stops the $\zeta$-recursion for $\zeta \geq \zeta_{max}$ where $\zeta_{max}$ is generally chosen in a problem-specific manner.

Given the additive nature of the gain-based updates yielding un-weighted sequence of particle systems converging in measure to the solution of the KS equation, one may draw a direct analogy of the form of the KS equation with the ensemble variants of the KF (herein generically referred to as the EnKF), even as most of these methods are limited to Gaussian approximations to the filtering PDF. Nevertheless, since the EnKF provides a ready framework that combines Monte Carlo simulations with KF-like additive updates, it may be prudent to borrow part of the algorithmic setup/terminology whilst developing the iterative gain-based stochastic filter bank (IGSF Bank) scheme. This is accomplished by iteratively refining a gain matrix ($\mathcal{K}_i^l$, $l$ being the iteration index corresponding to an ordering $0 = \zeta_0 \leq \zeta_1 \leq ... \leq \zeta_{max}$ of the pseudo-time parameter) consistent with the additively split predicting and updating terms in the previously noted approximation of the KS equation. The IGSF-Bank is implemented here in two stages. An EnKF-like prediction-cum-update step forms the first stage which is followed by an iterative update stage employing an annealing-type ADP ($\alpha^l$, $l$ being the iteration index) and an iterated gain $\mathcal{K}_i^l$ within a Gaussian sum approximation framework.

### 3.1. Gaussian Sum Approximation and Filter Bank

Gaussian sum approximation makes use of a weighted sum of Gaussian densities (Gaussian mixture model) to approximate the posterior density, wherein a tractable solution is arrived at from the sufficient statistics [16]. Let $\mathcal{N}(^i\tilde{X};^iM,^i\Sigma)$ denote the normal density corresponding to the state vector $^i\tilde{X} := \tilde{X}(t); t \in (t_i\ t_{i+1}]$ with mean vector $^iM$ and covariance matrix $^i\Sigma$, assumed to be non-singular. However, since it suffices to elaborate the recursive filtering scheme only over the sub-interval $(t_i\ t_{i+1}]$, the left superscript '$i$' is removed in what follows. Then the following theorems from [22], reproduced here for completeness, elucidates the underlying principle.

***Lemma 1***: Any probability density $p(\tilde{X})$ can be approximated as closely as desired in the space $L_1(\mathbb{R}^J)$ by a weighted mixture of the form

$$p_{N_G}^{GS}(\tilde{X}) = \sum_{\eta=1}^{N_G} w^{(\eta)} \mathcal{N}(\tilde{X}, M^{(\eta)}, \Sigma^{(\eta)}) \tag{3.3}$$

for some integer $N_G$, positive scalars $w^{(\eta)}$ with $\sum_{\eta=1}^{N_G} w^{(\eta)} = 1$, mean vectors $M^{(\eta)}$ and positive definite matrices $\Sigma^{(\eta)}$, so that $\int_{\mathbb{R}^J} |p(x) - p_{N_G}^{GS}(x)|dx < \varepsilon$ for any given $\varepsilon > 0$.

***Theorem 1***: Suppose that the process and measurement noise vectors in equations (2.5 and 2.3b) are Gaussian with zero mean and covariances $\tilde{\mathscr{G}}$ and $\Sigma_Z := G_z G_z^T$ respectively. Denoting $\tilde{X}_{i+1} := \tilde{X}(t_{i+1})$, let the prediction density $p(\tilde{X}_{i+1} | Z_{1:i}) = \mathcal{N}(\tilde{X}_{i+1}; M_{i+1|i}, \Sigma_{i+1|i})$ be Gaussian, where the subscripts $i+1|i$ in $M_{i+1|i}, \Sigma_{i+1|i}$ indicate that the current measurement $Z_{i+1}$ is yet to be assimilated within these statistical quantities. Then for fixed $\mathcal{H}(.)$, $M_{i+1|i}$ and $\Sigma_Z$, the filtering density $p(\tilde{X}_{i+1} | Z_{1:i+1}) = c_{i+1} p(\tilde{X}_{i+1} | Z_{1:i}) p(Z_{i+1} | \tilde{X}_{i+1})$ converges uniformly to the Gaussian density $\mathcal{N}(\tilde{X}_{i+1}; M_{i+1|i+1}, \Sigma_{i+1|i+1})$ as $\Sigma_{i+1|i} \to 0$ (in terms of a suitable matrix norm or their traces). Here $c_{i+1}$ is the normalizing constant. Moreover, if the last filtering density $p(\tilde{X}_i | Z_{1:i}) = \mathcal{N}(\tilde{X}_i; M_{i|i}, \Sigma_{i|i})$ is Gaussian, then for fixed $\tilde{\mathcal{F}}(.), M_{i|i}$ and $\tilde{\mathscr{G}}$, the prediction density $p(\tilde{X}_{i+1} | Z_{1:i}) = \int p(\tilde{X}_{i+1} | \hat{\tilde{X}}_i) p(\hat{\tilde{X}}_i | Z_{1:i}) d\hat{\tilde{X}}_i$ converges uniformly to the Gaussian density $\mathcal{N}(\tilde{X}_{i+1}; M_{i+1|i}, \Sigma_{i+1|i})$ as $\Sigma_{i|i} \to 0$. In the above expressions, the subscripts $i+1|i+1$ denote quantities associated with the filtered solutions at time $t_{i+1}$. Instead of the PDFs, this theorem may be more generally stated in terms of the associated characteristic functions as well.

The theorem above (which can be proved using Bayes' formula aided by some algebraic manipulations [22]) implies that, with covariance matrices having relatively small norms, the filtering and prediction densities may be adequately approximated as Gaussian mixtures. Based on this approximation, one can construct a bank of $N_G$ Gaussian mixands wherein, at the start of recursion, i.e. at $t = 0$, equal number of particles are drawn from each Gaussian PDF in the mixand (i.e. the set of particles is split into $N_G$ subsets, each containing

$\gamma := N / N_G$ particles drawn from each mixand in the mixture) so as to populate the ensemble. This enables a tagging of subsets of particles with appropriate terms in the Gaussian sum all through the recursion/iteration stages. Also, the mixands are assigned equal weights $w_0^{(\eta)} := w^{(\eta)}(t_0) = \dfrac{1}{N_G}$ to begin with.

## 3.2 Prediction and the Zeroth Update

This stage of the algorithm consists of the conventional propagation and the initial (zeroth) update steps. Let $\{^{(\eta)}\hat{\tilde{X}}_{(u),i}\}_{u=1}^{\gamma}$ be the sub-ensemble, consisting of $\gamma$ realizations of the Gaussian random variable $^{(\eta)}\hat{\tilde{X}}_i$ associated with the $\eta^{th}$ mixand ($\eta \in [1, N_G]$) in the Gaussian sum approximation of the last filtering density at $t = t_i$. Let the sample mean and the sample covariance of this sub-ensemble be respectively denoted as $<^{(\eta)}\hat{\tilde{X}}_i>$ and $\hat{\Sigma}_i^{(\eta)}$. The numerical integration of the process SDEs may be accomplished through any available numerical/semi-analytical scheme, e.g. the locally transversal linearization (LTL) [23] or the phase space linearization (PSL) [24,25] etc. Specifically using the PSL, i.e. based on the linearized SDE 2.5, the predicted solution corresponding to the $\eta^{th}$ mixand, $^{(\eta)}\tilde{X}(t)$, $t \in (t_i, t_{i+1}]$, is:

$$^{(\eta)}\tilde{X}(t) = \exp(\tilde{\mathcal{Q}}(t - t_i))(^{(\eta)}\tilde{X}_i) + \exp(\tilde{\mathcal{Q}}(t - t_i))\int_{t_i}^{t}[\exp(-\tilde{\mathcal{Q}}(s - t_i))\tilde{G}(s)]d\tilde{B}(s) \quad (3.4)$$

The above solution at $t = t_{i+1}$ generates a sub-ensemble of predicted particles $\{^{(\eta)}\tilde{X}_{(u),i+1}, u \in [1, \gamma]\}$, whose sample mean vector is given by:

$$<^{(\eta)}\tilde{X}_{i+1}> = \dfrac{1}{\gamma}\sum_{u=1}^{\gamma}{}^{(\eta)}\tilde{X}_{(u),i+1} \quad (3.5)$$

To elaborate the zeroth update scheme, the so called prediction error (anomaly) and predicted measurement anomaly matrices are respectively defined as:

$$S_{i+1}^{(\eta)} = \dfrac{1}{\sqrt{\gamma - 1}}[(^{(\eta)}\tilde{X}_{(1),i+1} - <^{(\eta)}\tilde{X}_{i+1}>), \ldots, (^{(\eta)}\tilde{X}_{(\gamma),i+1} - <^{(\eta)}\tilde{X}_{i+1}>)] \quad (3.6)$$

$$(S_z^{(\eta)})_{i+1} = \dfrac{1}{\sqrt{\gamma - 1}}[(\mathcal{H}(^{(\eta)}\tilde{X}_{(1),i+1}) - <\mathcal{H}(^{(\eta)}\tilde{X}_{i+1})>), \ldots, (\mathcal{H}(^{(\eta)}\tilde{X}_{(\gamma),i+1}) - <\mathcal{H}(^{(\eta)}\tilde{X}_{i+1})>)] \quad (3.7)$$

The above are used in generating the zeroth iterate, which is input to the iterative updating procedure outlined in the next section. The zeroth update (filtering) step, fashioned after the currently adopted particle-based form of the EnKF, takes in the current observation and generates the updated particles at $t = t_{i+1}$ via:

$$^{(\eta)}\hat{\tilde{X}}^0_{(u),i+1} = {}^{(\eta)}\tilde{X}_{(u),i+1} + \mathcal{K}^{(\eta),0}_{i+1}(Z_{i+1} - \mathcal{H}({}^{(\eta)}\tilde{X}_{(u),i+1})), \ u = [1,\gamma], \ \eta = [1, N_G] \quad (3.8)$$

Here $\mathcal{K}^{(\eta),0}_{i+1}$ is the initial gain matrix (with the superscript '0' indicating the zeroth update) defined through:

$$\mathcal{K}^{(\eta),0}_{i+1} = S^{(\eta)}_{i+1}(S^{(\eta)}_z)^T_{i+1}((S^{(\eta)}_z)_{i+1}(S^{(\eta)}_z)^T_{i+1} + \Sigma_Z)^{-1} \quad (3.9)$$

Thus, if $\pi_{i+1,\zeta_0}(\phi({}^{(\eta)}\hat{\tilde{X}})) := \pi^0_{i+1}(\phi({}^{(\eta)}\hat{\tilde{X}}))$ denotes the mean of the $\eta^{th}$ filtered PDF component in the Gaussian sum following the zeroth iterate as above, then its sample approximation, upon averaging over the associated sub-ensemble with $\gamma$ elements, is given by $\pi^0_{i+1}(\phi({}^{(\eta)}\hat{\tilde{X}})) \cong < \phi({}^{(\eta)}\hat{\tilde{X}}^0_{(u),i+1}) >$.

## 3.3 The ADP-based Iterative Update Scheme

The second stage involves repetitive updates over the zeroth filtered particles via iterations. The iterative updates entail, for every given time $i+1$ (i.e. $t_{i+1}$) an iterating index $l = 1, 2, ...$ that may be construed as being correspondent to a discretely introduced pseudo time parameter set $\{\zeta_l : \zeta_1 < ... < \zeta_\Gamma = \zeta_{max}\}$, with $\zeta_0 = 0$ (corresponding to the zeroth update) and $\Gamma$ denoting the maximum number of iterations. Let $\pi_{t,\zeta_l}(.) := \pi^l_t(.)$ denote the estimate at the end of the $l^{th}$ iterate. Then the iterative updates (averaged over process noise) over $(t_i, t_{i+1}]$ may be written as:

$$\pi^l_{i+1}(\phi({}^{(\eta)}\hat{\tilde{X}})) = \mathrm{E}_\mathrm{P}(\Psi_i(\phi({}^{(\eta)}\tilde{X}))) + (1+\alpha^l)\mathcal{K}^{(\eta),l-1}_{i+1}(Z_{i+1} - \mathcal{H}(\{\pi^{l-1}_{i+1}({}^{(\eta)}\hat{\tilde{X}}^{(j)}) : j \in [1, J]\})) \quad (3.10a)$$

where $\Psi_i$ stands for an explicit time marching map (as obtainable from the locally linearized solution Eq. 3.4) upon integrating the process SDE such that $\phi({}^{(\eta)}\tilde{X}_{i+1}) = \Psi_i(\phi({}^{(\eta)}\tilde{X})) = \Psi(\phi({}^{(\eta)}\tilde{X}_i))$. Moreover $\mathcal{K}^{(\eta),l-1}_{i+1}$ denotes the $l-1^{th}$ iterative update of the gain matrix corresponding to the $\eta^{th}$ mixand in the Gaussian sum and $\alpha^l := \alpha(\zeta_l)$ is the pseudo-time dependent ADP that can be likened to the annealing parameter typically used with simulated annealing (SA) applied to a Markov chain [26]. Unlike the SA where the temperature is recursively reduced to zero whilst evolving a single Markov chain, the sequence $\{\alpha^l \mid \alpha^{l+1} \leq \alpha^l\}$ is used in the present filter to evolve an ensemble of $N = \gamma N_G$ pseudo-Markov chains in $\zeta$ so that, for a given $t$, the chains proceed in a controlled way to finally arrive at an ensemble that drives the pseudo innovation process $I_{t,\zeta}$ to a zero-mean Brownian motion in $\zeta$. Analogous to the SA, one must then supplement the iterative scheme with $\alpha^1$, the initial ADP, and an appropriate schedule to drive $\alpha^l$ to zero over successive iterations. However, given that a Monte Carlo scheme (based on an ensemble of approximations to $\hat{\tilde{X}}_{t,\zeta}$) is in itself a means to efficaciously explore the phase space, the conservative or even geometric annealing schedules, typically used with standard SA algorithms or MCMC filters [27] and involving a large number of iterations, need not be adhered to here. Indeed, as the numerical experiments confirm, considerable flexibility with the scheduling of the ADP (as well as in the choice of $\alpha^1$) is possible with the proposed filter. Specifically, the exponentially decaying schedule adopted here is given by $\alpha^{l+1} := \alpha(\zeta_{l+1}) = \dfrac{\alpha^l}{\exp(l)}$, with $\alpha^1$ determined in a problem-specific manner through a few trial runs (alternatively, it may be computed as the empirical average of an instantaneously defined cost functional evaluated at the particle locations corresponding to the zeroth update). The scheduled sequence of the ADP $\{\alpha^l\}$ is intended to provide an additional handle in controlling the mixing property of the iterative update kernels and ensure that the process variables visit every finite subset of the phase space of interest sufficiently frequently [17]. Finally, the uncoupled nature of the prediction and iterative update may be observed from the fact that the latter affects only the third term on the RHS of the KS Equation (3.1) whilst leaving unaltered the prediction part of the estimate (i.e. the first two terms on the RHS of Equation 3.1).

Replacing the expectation operators appearing in $\pi_{i+1}^l, \pi_{i+1}^{l-1}$ and $E_P$ by appropriate sub-ensemble averages, Eq. (3.10a) may be modified to go with MC simulations involving finite ensembles as:

$$<\phi(^{(\eta)}\hat{\tilde{X}}^l_{(u),i+1})> = <\Psi_i(\phi(^{(\eta)}\tilde{X}_{(u)}))> + (1+\alpha^l)\mathcal{K}^{(\eta),l-1}_{i+1}(Z_{i+1} - <\mathcal{H}(^{(\eta)}\hat{\tilde{X}}^{l-1}_{(u),i+1})>) \qquad (3.10b)$$

A particle based version of the above, as implemented in this work, becomes:

$$\phi(^{(\eta)}\hat{\tilde{X}}^l_{(u),i+1}) = \Psi_i(\phi(^{(\eta)}\tilde{X}_{(u)})) + (1+\alpha^l)\mathcal{K}^{(\eta),l-1}_{i+1}(Z_{i+1} - \mathcal{H}(^{(\eta)}\hat{\tilde{X}}^{l-1}_{(u),i+1})); \; u \in [1,\gamma] \qquad (3.11)$$

with $\{^{(\eta)}\hat{\tilde{X}}^l_{(u),i+1}\}$ denoting the ensemble of filtered particles $t=t_{i+1}$ following the $l^{th}$ update. The updated prediction and measurement anomaly matrices defined respectively as

$$\hat{S}^{(\eta),l-1}_{i+1} = \frac{1}{\sqrt{\gamma-1}}[(^{(\eta)}\hat{\tilde{X}}^{l-1}_{(1),i+1} - {}^{(\eta)}\tilde{X}_{(1),i+1}),...,(^{(\eta)}\hat{\tilde{X}}^{l-1}_{(\gamma),i+1} - {}^{(\eta)}\tilde{X}_{(\gamma),i+1})] \quad (3.12)$$

$$(\hat{S}^{(\eta),l-1}_z)_{i+1} = \frac{1}{\sqrt{\gamma-1}}[(\mathcal{H}(^{(\eta)}\hat{\tilde{X}}^{l-1}_{(1),i+1}) - Z_{i+1}),...,(\mathcal{H}(^{(\eta)}\hat{\tilde{X}}^{l-1}_{(\gamma),i+1}) - Z_{i+1})] \quad (3.13)$$

are then used for evaluating the updated gain matrix $\mathcal{K}^{(\eta),l-1}_{i+1}$ as:

$$\mathcal{K}^{(\eta),l-1}_{i+1} = \hat{S}^{(\eta),l-1}_{i+1}(\hat{S}^{(\eta),l-1}_z)^T_{i+1} ((\hat{S}^{(\eta),l-1}_z)_{i+1}(\hat{S}^{(\eta),l-1}_z)^T_{i+1})^{-1} \quad (3.14)$$

where the conventions $\hat{S}^{(\eta),0}_{i+1} := S^{(\eta)}_{i+1}$ and $(\hat{S}^{(\eta),0}_z)_{i+1} := (S^{(\eta)}_z)_{i+1}$ are adopted. The mixand weights are then updated and normalized using the particles $\{^{(\eta)}\hat{\tilde{X}}^\Gamma_{(u),i+1}\}$, available after the last (i.e. $\Gamma^{th}$) iteration, as:

$$\tilde{w}^{(\eta)}_{i+1} = w^{(\eta)}_{i+1} \frac{\mathcal{N}(Z_{i+1};<\mathcal{H}(^{(\eta)}\hat{\tilde{X}}^\Gamma_{i+1})>,(\hat{S}^{(\eta),\Gamma}_z)_{i+1}(\hat{S}^{(\eta),\Gamma}_z)^T_{i+1} + \Sigma_Z)}{\sum_{\eta=1}^{N_G}\mathcal{N}(Z_{i+1};<\mathcal{H}(^{(\eta)}\hat{\tilde{X}}^\Gamma_{i+1})>,(\hat{S}^{(\eta),\Gamma}_z)_{i+1}(\hat{S}^{(\eta),\Gamma}_z)^T_{i+1} + \Sigma_Z)}, \eta=1,...,N_G \quad (3.15)$$

followed by $w^{(\eta)}_{i+1} = \frac{\tilde{w}^{(\eta)}_{i+1}}{\sum_{\eta=1}^{N_G}\tilde{w}^{(\eta)}_{i+1}}$ \quad (3.16)

Also, by convention, the last updated particles corresponding to the $\eta^{th}$ mixand in the Gaussian sum are denoted as $\{^{(\eta)}\hat{\tilde{X}}^\Gamma_{(u),i+1}\} := \{^{(\eta)}\hat{\tilde{X}}_{(u),i+1}\}$ before carrying them over to the next time step. Statistics estimation can be done based on the empirical posterior probability density function so obtained. Specifically, the (sample) estimate of the state vector $\tilde{X}$ at $t=t_{i+1}$ is given by:

$$<\hat{\tilde{X}}_{i+1}> = \sum_{\eta=1}^{N_G} w^{(\eta)}_{i+1} <^{(\eta)}\hat{\tilde{X}}_{i+1}> \quad (3.17)$$

This filter bank reduces to a single filter (IGSF with ADP) when $\eta=1$.

## 4. Numerical Illustrations

For state estimation, the performance of the proposed filter is illustrated via a 1-dimensional nonlinear system and a target tracking problem. Towards assessing the performance of the filter for combined state-parameter estimation and by way of highlighting its efficacy in resolving higher dimensional nonlinear filtering problems, a multi-degree-of-freedom (MDOF) shear frame model is adopted.

### 4.1. 1-Dimensional Nonlinear System with Additive Gaussian Noise

A univariate nonstationary growth model with additive Gaussian noise is chosen as the first numerical example, wherein both the process and measurement equations are nonlinear. Similar examples have been widely used [5, 28, 29, 30] for the performance evaluation of various particle filters. A discrete version of the governing equation of the nonlinear system (or, equivalently, the predicted solution) considered here may be written as:

$$X_{i+1} = (\gamma_1 X_i + \gamma_2 X_i^2 + 8\cos(\vartheta i))h + G \Delta B_i \tag{4.1}$$

where $\gamma_1$, $\gamma_2$ and $\vartheta$ are scalar system parameters. Here $h$, the time-step size, is chosen as 1s and the reference parameter values are given by $\gamma_1 = 0.2$, $\gamma_2 = 0.01$ and $\vartheta = 1.2$. Process noise variance is $G^2$ and $\Delta B_i := B(t_{i+1}) - B(t_i)$ is the Brownian increment over $h$ where $B(t)$ denotes a standard Brownian noise starting at zero. The state is estimated from the measurement equation given by

$$Z_{i+1} = X_{i+1}^2 + G_z (\Delta B_z)_i \tag{4.2}$$

where the measurement noise variance $G_z^2$ and $B_z(t)$, another standard Brownian motion. The performance of the IGSF Bank (always ADP-based, unless explicitly noted otherwise) is compared with the Gaussian sum particle filter (GSPF) [31] for state estimation. The root-mean-squared-error (RMSE) via the IGSF Bank and GSPF over 100 independent Monte Carlo runs are computed using an ensemble size $N = 1000$ and with $N_G = 10$. Here, the assumptions as in [28] are followed in that the reference initial state is assumed to be a uniformly distributed random variable in [0, 1] and prior state at $t_0 = 0$ is taken as $\tilde{X}_0 = \hat{\tilde{X}}_0 \sim \mathcal{N}(0.5, 2)$. The initiating ADP $\alpha^1$ and the number of updating iterations $\Gamma$ are chosen as 1 and 5 respectively. Figs 1(a) and (b) show the time histories of estimate RMSEs by the two filters for $G_z^2 = 0.01$ and 10 respectively. Both use a constant process noise variance $G^2 = 10$.

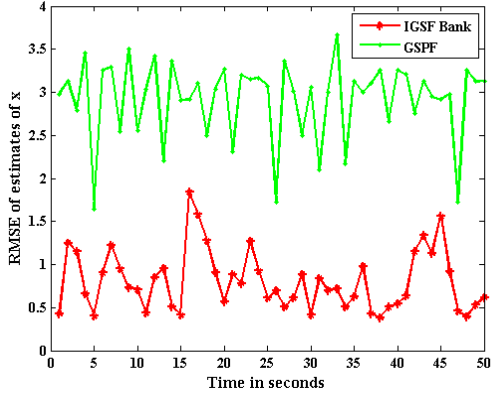 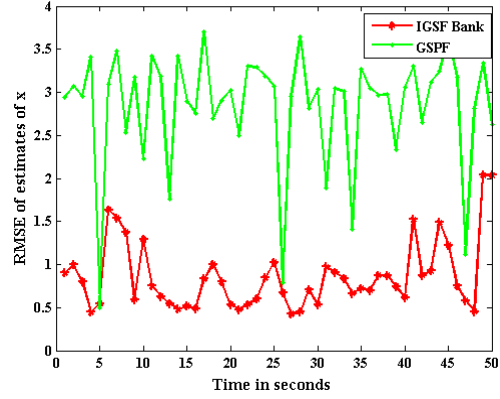

(a)                        (b)

Fig 1. Time histories of RMSE of the estimates via IGSF Bank and GSPF: (a) $G^2 = 0.01$ and (b) $G^2 = 10$.

The observed performance of the IGSF Bank, as evidenced from the results in Fig 1, is indicative of consistently improved estimation accuracy irrespective of the measurement noise level when compared to the GSPF.

### 4.2. Target Tracking

In a target tracking problem, one typically estimates the trajectory of a maneuvering target (i.e. position and velocity) from the noise-corrupted sensor bearing and range data. The dynamic model of the maneuvering target (not the process model, rather the one used to synthetically generate the measurement) adopted here in a discretized form is:

$$\Xi_{i+1} = \Upsilon \Xi_i + \Lambda [a_i + m_i] \tag{4.3}$$

where $\Upsilon = \begin{bmatrix} 1 & \Delta & 0 & 0 \\ 0 & 1 & 0 & 0 \\ 0 & 0 & 1 & \Delta \\ 0 & 0 & 0 & 1 \end{bmatrix}$ and $\Lambda = \begin{bmatrix} \frac{1}{2}\Delta^2 & 0 \\ \Delta & 0 \\ 0 & \frac{1}{2}\Delta^2 \\ 0 & \Delta \end{bmatrix}$

$\Xi_i = \begin{bmatrix} X & X_v & Y & Y_v \end{bmatrix}_i^T$ is the state vector with $X$ and $X_v$ respectively being the position and velocity of the moving target along the Cartesian space $x$-axis and $Y, Y_v$ denoting similar variables along the $y$-axis. Moreover,

$$m_i = \begin{cases} \{\varpi_1, \varpi_2\}_i^T & \text{if } t_i = s_\varpi \\ \mathbf{0} & \text{otherwise} \end{cases}$$

is the acceleration vector that brings in manoeuvring at the chosen time instants $s_\varpi \in [0, t_1, ..., t_i, ..., \tau]$, $a_i$ is the random acceleration vector of the target, currently characterized as Brownian, $\Delta$ is the constant sampling interval (state update period) and $\mathbf{0} \in \mathbb{R}^2$. The sensor is situated at the origin $(x_0, y_0)$ of the plane with the bearing angle and the distance from the sensor taken as the measurements according to the measurement equation:

$$Z_{i+1} = \begin{bmatrix} \tan^{-1}\left(\frac{Y_{i+1}-Y_0}{X_{i+1}-X_0}\right) \\ \sqrt{(Y_{i+1}-Y_0)^2 + (X_{i+1}-X_0)^2} \end{bmatrix} + v_{i+1} \qquad (4.4)$$

$v_{i+1} = \{v_{z1}, v_{z2}\}_{i+1}^T$ is a zero mean white Gaussian sequence with covariance $G_z G_z^T = \text{diag}([(G_z)_{11}^2 \ (G_z)_{22}^2])$. The actual initial state of the target (based on which the noise-corrupted synthetic measurement is generated) is chosen to be at [0.5m 3ms$^{-1}$ 1m 1ms$^{-1}$] in the Cartesian coordinates. From here it undergoes 3-leg manoeuvring sequences by taking sharp turns at 20s, 30s and 60s with respective accelerations [-40ms$^{-2}$ 40 ms$^{-2}$], [25 ms$^{-2}$ -25 ms$^{-2}$] and [25 ms$^{-2}$ -25 ms$^{-2}$] whilst moving along straight lines with constant velocities for the intervals in between till the trajectory ends at $\tau = 80$s. Assuming that the dynamic model (4.3) of the manoeuvring target is unknown, we consider a simpler motion model $\Xi_{i+1} = \Upsilon \Xi_i + \Lambda w_i$ for tracking, where $w_i$, the random acceleration of the target, is a zero-mean Gaussian process noise sequence with covariance chosen as diag([8m$^2$s$^{-4}$ 8 m$^2$s$^{-4}$]). For initiating the filter, the prior state at $t_0 = 0$ is taken as Gaussian with the mean vector [0m 40ms$^{-1}$ 0.2m 0.075ms$^{-1}$] (which is far away from the true state) and sampling interval $\Delta$ set to 0.1s. The measurement noise covariance is chosen as diag([0.2rad$^2$ 35m$^2$]). Filter parameters for the IGSF Bank are chosen as $\alpha^1 = 10$, $\Gamma = 10$ and $N_G = 5$. The performance of the proposed filter with an ensemble size of $N = 200$ particles is compared with that of auxiliary sampling importance resampling filter (ASIR) [32] in Fig. 2, which reports the estimated tracks of the target states.

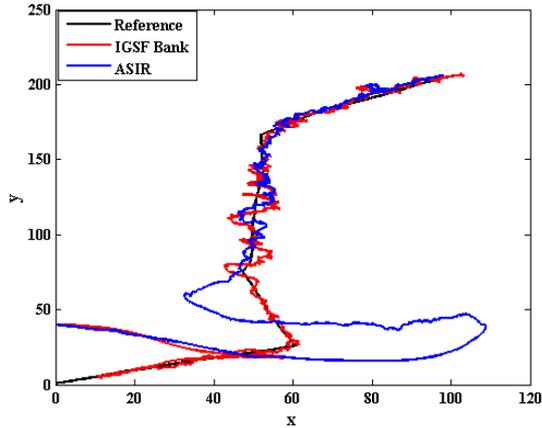

Fig 2. Estimated target tracks in the *x-y* plane by the IGSF Bank and the ASIR along with the true trajectory as reference

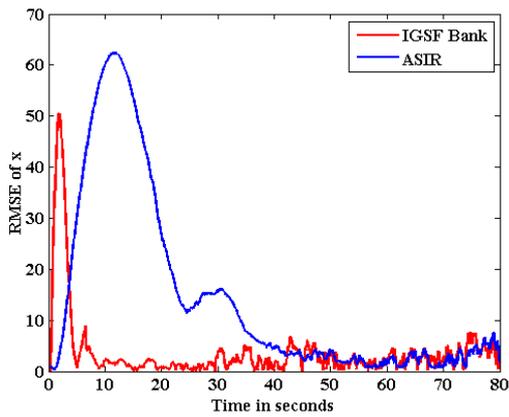  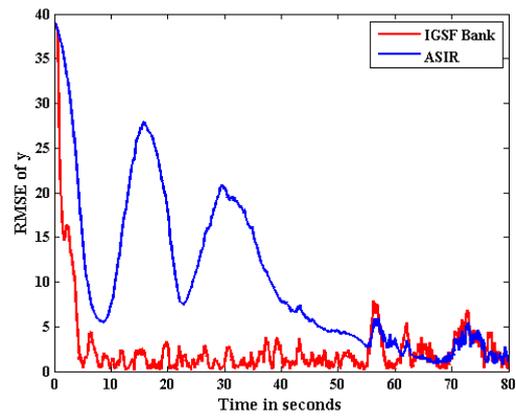

(a) (b)

Fig 3. Evolutions of the RMSE of the (a) *x*-coordinate and (b) *y*-coordinate of the tracked target via the IGSF Bank and the ASIR

A comparison of the RMSEs of the estimated *x* and *y* positions of the target is given in Fig 3. The vastly improved performance of the IGSF bank over the ASIR in terms of faster convergence and estimation accuracy is evident, despite sampling fluctuations inevitable with a small ensemble size. As expected, with a larger ensemble size the performance of ASIR also improves.

### 4.3. An MDOF Shear Frame Model

An MDOF shear frame model, schematically depicted in Fig 4, is chosen to evaluate the performance of the proposed filter for higher dimensional state-cum-parameter identification problems. The governing differential equation of the model with an additive Brownian diffusion term is formally represented as:

$$\ddot{X} + C\dot{X} + SX = F(t) + G\dot{B}(t) \qquad (4.5)$$

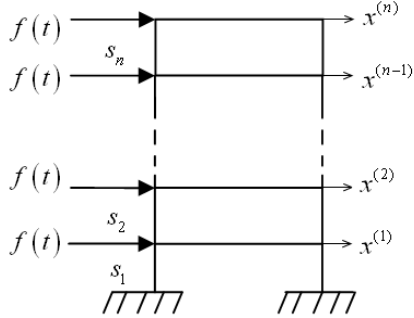

Fig 4. An $n$-DOF shear frame model

Here $S$ and $C$, respectively denoting the $n \times n$ stiffness and viscous damping matrices, are of the form:

$$S = \begin{bmatrix} s_1 + s_2 & -s_2 & 0 & 0 & 0 \\ -s_2 & s_2 + s_3 & -s_3 & 0 & 0 \\ 0 & ... & .. & ... & 0 \\ 0 & 0 & -s_{n-1} & s_{n-1} + s_n & -s_n \\ 0 & 0 & 0 & -s_n & s_n \end{bmatrix} \qquad (4.6)$$

$$C = \begin{bmatrix} c_1 + c_2 & -c_2 & 0 & 0 & 0 \\ -c_2 & c_2 + c_3 & -c_3 & 0 & 0 \\ 0 & ... & .. & ... & 0 \\ 0 & 0 & -c_{n-1} & c_{n-1} + c_n & -c_n \\ 0 & 0 & 0 & -c_n & c_n \end{bmatrix} \qquad (4.7)$$

A deterministic force vector $F(t) = \{f^{(j)}(t)\} \in \mathbb{R}^n$, with $f^{(j)}(t) = f_0 \cos(\vartheta_f t) \; \forall j \in [1, n]$, is applied at each degree-of-freedom. Also, the noise intensity matrix $G$ is presently an $n \times n$ diagonal matrix. The state vector is $\bar{X} = \left[ X^{(1)}, \dot{X}^{(1)}, X^{(2)}, \dot{X}^{(2)}, ..., X^{(n)}, \dot{X}^{(n)} \right]^T$ and $\bar{X}_0 = \mathbf{0} \in \mathbb{R}^{2n}$. Note that the combined state-parameter estimation problem is here a nonlinear filtering problem, even though the process dynamics would be strictly linear if the parameters were known. Including the unknown (stiffness and damping) parameters as additional states, the augmented process state vector is given by $\tilde{X} := [\bar{X}^T, \mu^T]^T, \tilde{X} \in \mathbb{R}^{4n}$, with $n_\mu = 2n$ denoting the parameter dimension and $J = 4n$. Given the ability of proposed filters (the IGSF Bank as well as the ADP-based IGSF) to work with low measurement noise levels (possible with sophisticated measuring devices), the data (herein consisting only of the noisy displacement DOFs) is synthetically generated by adding a very low noise intensity (less than 1%) to all the elements of displacement vector $X$. Particle filters [33] typically perform poorly with very low-intensity measurement noises, employed here to improve the estimation quality (i.e. to avoid random oscillations owing to larger variance in the measurement noise).

The performance of the IGSF Bank is compared with the EnKF (which is known to tackle relatively higher dimensional filtering problems), IGSF Bank and IGSF with ADP (the degenerate case of the IGSF bank with $N_G = 1$) for a 20-dimensional (5-DOF) system with stiffness and damping parameters respectively chosen as 100N/m and 5Ns/m (uniformly) along with a forcing amplitude 30N and frequency 5 rad/s. While the ensemble size is consistently given by $N = 400$, the other algorithmic parameters of relevance are fixed as $N_G = 10$, $\Gamma = 10$ and $\alpha^1 = 2$. In order to demonstrate the possible flexibility in scheduling the sequence $\{\alpha^l\}$, $\alpha^1$ for this example is not exponentially decreased as in the previous examples, but kept constant till the $(\Gamma-1)^{\text{th}}$ iterative update and discontinuously reduced to zero in the last iterate. The estimates of stiffness parameters via EnKF, IGSF, IGSF with ADP and the IGSF Bank are shown in Fig 5. The superior performance features of the last two filters are thus clearly brought forth.

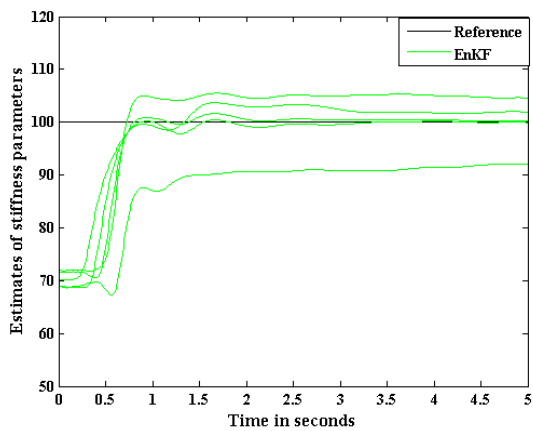

(a)

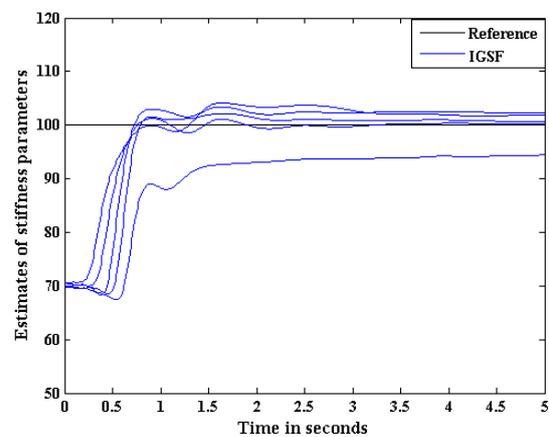

(b)

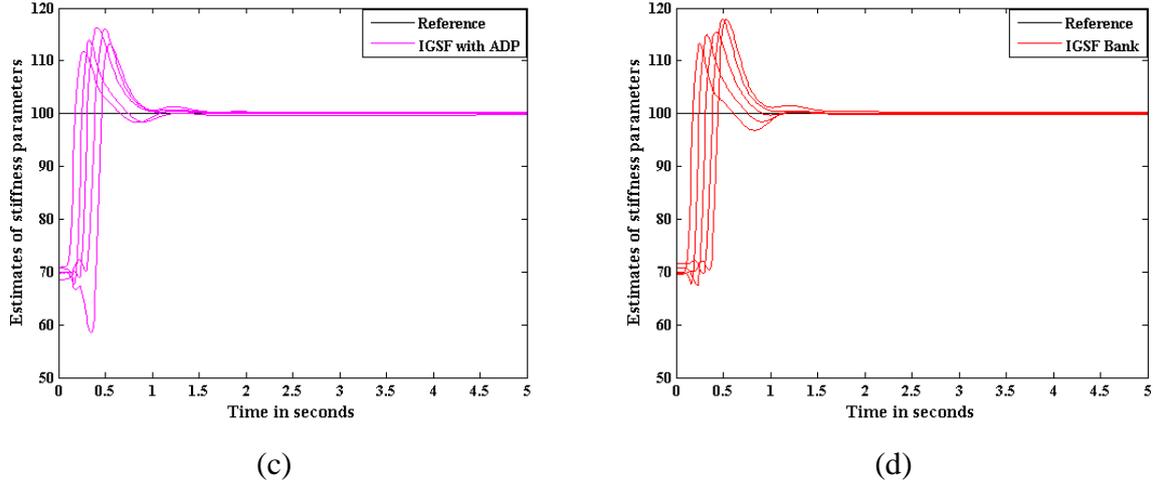

| (c) | (d) |

Fig 5. Estimates of stiffness parameters for the 5-DOF shear frame model via (a) EnKF, (b) IGSF, (c) IGSF with ADP and (d) IGSF Bank

The contrast in the performance the proposed filters vis-à-vis a few existing ones may be further highlighted by considering the combined state-parameter estimation of a 20-DOF shear frame ($n=20$), yielding an 80-dimensional ($J=80$) nonlinear filtering problem. In order to verify if the new filtering schemes can successfully treat 'incipient' damage/degradation scenarios often characterized by small local changes in (some of) the parameter profiles, the stiffness parameters $s_{19}$ and $s_{20}$ in the reference shear frame model (whose response, following corruption by low intensity noise, provides the synthetic measurement) are taken as 98 N/m with $s_1$ through $s_{18}$ remaining 100 N/m as in the 5-DOF example. While all other model/numerical parameters are also kept the same as in the last example, $\alpha^1 = 3$ (slightly higher with respect to the 5-DOF case owing to the measurement noise intensity being lower) and $\Gamma = 8$ are presently used. In addition to the higher dimensional nature of the problem, wherein particle filters often fail to work, low-intensity measurement noises contribute to an additional performance barrier, needed nevertheless if small variations in the estimates were to be detected. Performance of the IGSF Bank is compared with the EnKF, IGSF and IGSF with ADP. Specifically, the estimates of stiffness and damping parameters via all these filters are shown in Figs 6 and 7 respectively. As observed in Fig 6, while the IGSF with ADP continues to perform better than the EnKF and the IGSF, only the IGSF Bank appears to resolve the problem with a good measure of success. Fig 7 also reveals a substantively superior resolution of the damping parameters, probably being reported for the first time to the authors' knowledge, through the IGSF Bank.

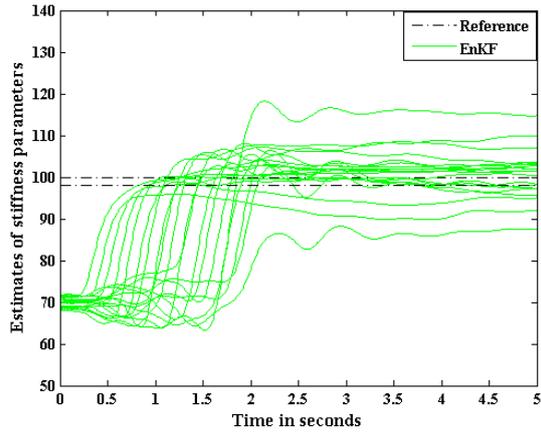
(a)

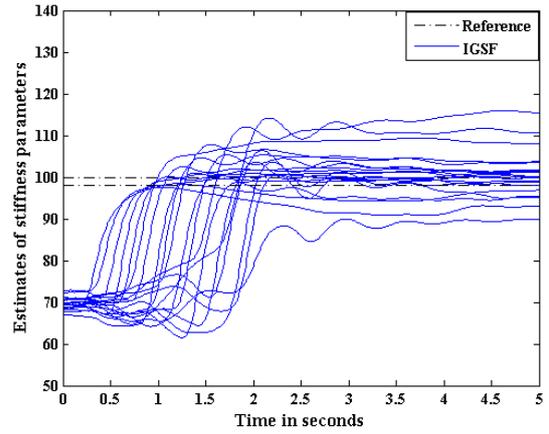
(b)

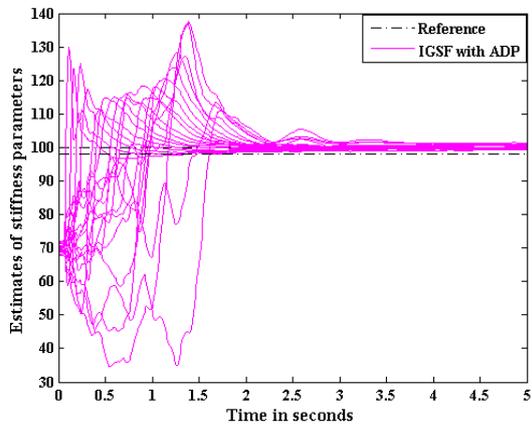
(c)

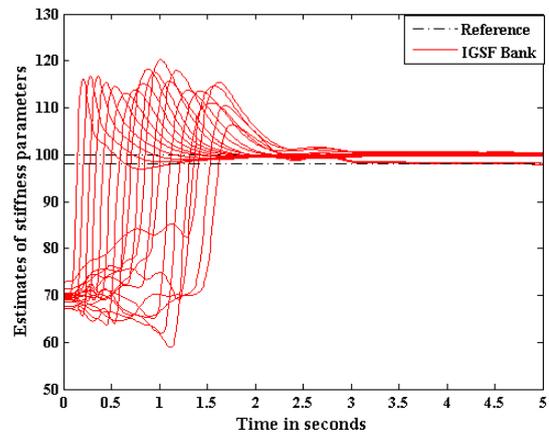
(d)

Fig 6. Estimates of stiffness parameters for a 20-DOF shear frame model via (a) EnKF, (b) IGSF, (c) IGSF with ADP and (d) IGSF Bank

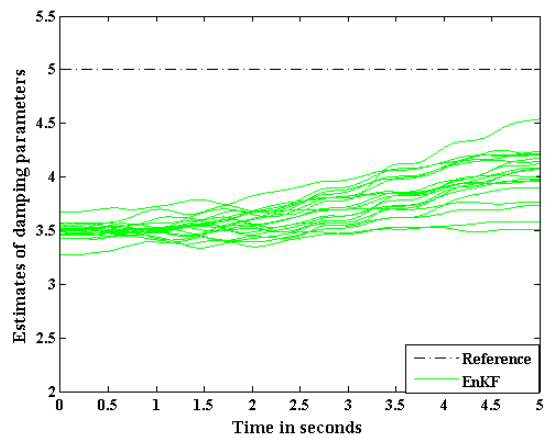
(a)

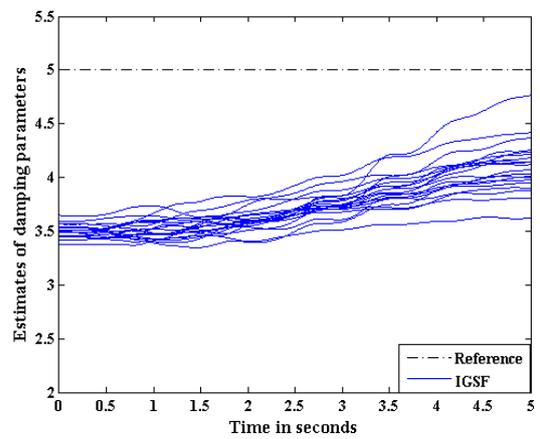
(b)

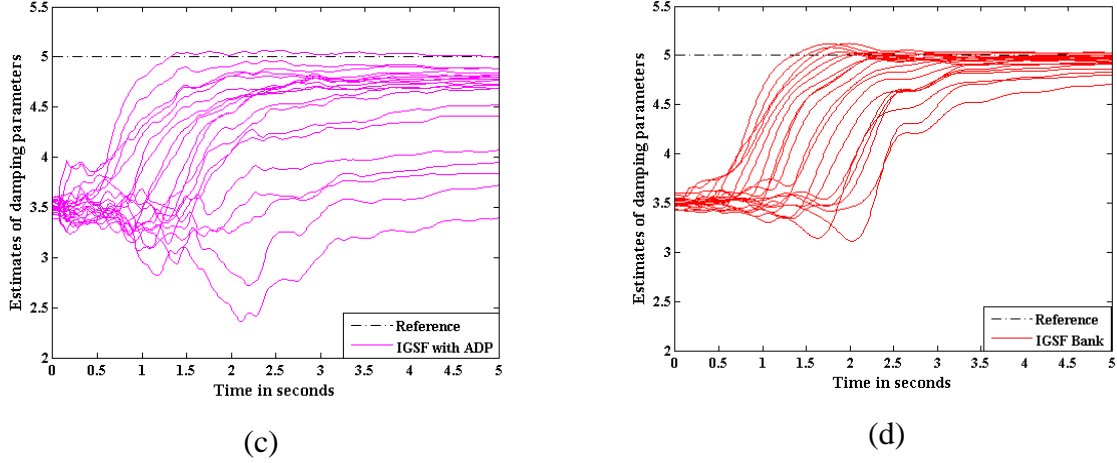

Fig 7. Estimates of damping parameters for an 20-DOF shear frame model by (a) EnKF, (b) IGSF, (c) IGSF with ADP and (d) IGSF Bank

## 5. Conclusions

Motivated through a discretized and iterative approximation to the Kushner-Stratonovich nonlinear filtering equations, a recursive-iterative Monte Carlo filtering approach employing a Gaussian sum approximation to the filtering density is proposed in this work. Employing additive gain-based iterated updates whilst assimilating the current measurement within the predicted solution, the proposed particle filters also make use of an additional annealing-type scalar parameter in order to boost the diffusion provided by the innovation term. However, unlike simulated annealing or MCMC filters using an annealing parameter, the value of the parameter at the beginning of iterative updates as well as the schedule used to bring it down to zero by the end of iterations may be chosen much more flexibly in the present filtering schemes. For instance, schedules with steep exponential and even discontinuous decay, thereby implying far less iterations, are admissible without requiring any burn-in periods. Whilst keeping the predicted particle locations unchanged at a given time, the iterative updates aim at so refining the particle locations as to drive the so-called measurement error to a zero-mean Brownian motion in a pseudo-time variable with respect to which the iterations may be parametrized. Using the proposed filters, non-trivial improvement in filter convergence as well as estimation accuracy (involving lower sampling variance) is observed consistently across all the numerical examples considered here on nonlinear state estimation and/or system identification.